\documentclass[letterpaper]{jpconf}
\usepackage{graphicx}
\usepackage[square,sort&compress]{natbib}
\begin{document}
\title{Evolution of single-ion crystal field and Kondo features in Ce$_{0.5}$La$_{0.5}$Ni$_{9-x}$Cu$_x$Ge$_4$}

\author{L Peyker$^1$, C Gold$^1$, E-W Scheidt$^1$, H Michor$^2$ and W Scherer$^1$}

\address{$^1$ CPM, Institut f\"{u}r Physik, Universit\"{a}t Augsburg, 86159
Augsburg, Germany}
\address{$^2$ Institut f\"{u}r Festk\"{o}rperphysik, Technische Universit\"{a}t Wien,
1040 Wien, Austria}

\ead{Ernst-Wilhelm.Scheidt@physik.uni-augsburg.de}

\begin{abstract}
Starting with the heavy fermion compound CeNi$_9$Ge$_4$, the
substitution of nickel by copper leads to a dominance of the RKKY
interaction in competition with the Kondo and crystal field interaction.
Consequently, this results in an antiferromagnetic phase transition in
CeNi$_{9-x}$Cu$_x$Ge$_4$ for $x>0.4$, which is, however, not fully completed
up to a Cu-concentration of $x=1$. To study the influence of single-ion
effects on the AFM ordering by shielding the $4f$-moments, we
analyzed the spin diluted substitution series
La$_{0.5}$Ce$_{0.5}$Ni$_{9-x}$Cu$_x$Ge$_4$ by
magnetic susceptibility $\chi$ and specific heat $C$ measurements. For small
Cu-amounts $x\leq 0.4$ the data reveal single-ion scaling with regard to the
Ce-concentration, while for larger Cu-concentrations the AFM
transition (encountered in the CeNi$_{9-x}$Cu$_x$Ge$_4$ series) is found to be
completely depressed. Calculation of the entropy reveal that the Kondo-effect
still shields the 4$f$-moments of the Ce$^{3+}$-ions in CeNi$_8$CuGe$_4$.
\end{abstract}

\section{Introduction}
One of the most outstanding Fermi-liquid systems is the heavy
fermion compound CeNi$_9$Ge$_4$, which turns out to display the largest
ever recorded value of the electronic specific heat $\Delta
C/T\approx5.5$\,Jmol$^{-1}$K$^{-2}$ without showing any magnetic order
\cite{Michor2004}. The dilution of the magnetic Cerium 4$f$-moments
in Ce$_{1-x}$La$_x$Ni$_9$Ge$_4$ reveals single-ion scaling with regard to the
Ce-concentration \cite{Killer2004}. Therefore, the unique behavior of
CeNi$_9$Ge$_4$ could be mainly attributed to a single-ion effect.
Gradually replacing Ni by Cu changes both, the 3$d$ electron number and
the lattice parameters. This substitution influences the crystal field and leads
to a formation of long range antiferromagnetic order in
CeNi$_{9-x}$Cu$_x$Ge$_4$ for $x>0.4$ which culminates in a transition temperature
of $T_{\rm{N}}=175$\,mK for $x=1$ \cite{Peyker2009}.
Even though the maximum of the magnetic specific heat
$\Delta C(T_{\rm{N}})$ of CeNi$_{8}$CuGe$_4$ reaches less than 15\%
of the theoretical expected value, the transition was discussed in terms of a
reduced long range antiferromagnetic order due to the presence of
the Kondo-effect \cite{Peyker2009}. At a suitable concentration of $x \simeq 0.4$, where a
crossover between single-ion and magnetic ordered behavior occurs, the system
exhibits a quantum critical phase transition (QCP)\cite{Peyker2009}.\\
In the present work we studied the influence of Kondo-shielding
on the antiferromagnetic ordering and how far single-ion effects
are still present when crossing the phase transition from a Kondo-state
($x \leq 0.4$) to an antiferromagnetic coherent state ($x \geq 0.4.$).
Therefore we performed magnetic susceptibility $\chi$
and specific heat $C$ measurements of the  spin diluted substitution series
La$_{0.5}$Ce$_{0.5}$Ni$_{9-x}$Cu$_x$Ge$_4$ and compared them to the
pure CeNi$_{9-x}$Cu$_x$Ge$_4$ series.

\section{Experimental Details}
Polycrystalline samples of
\begin{figure}
\begin{center}
\includegraphics[width=22pc]{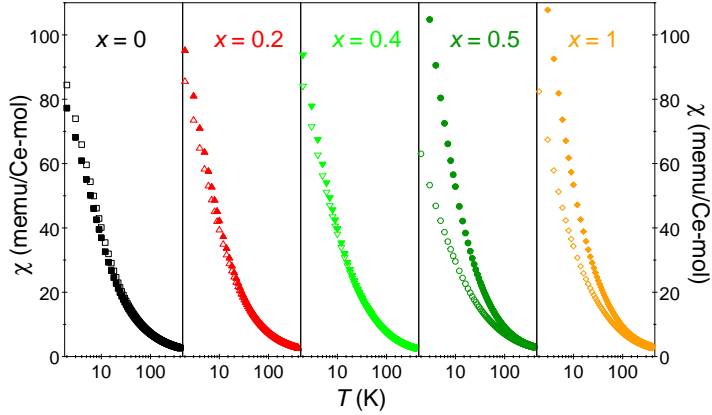}
\caption{\label{fig1}(Color online) The magnetic susceptibility
$\chi$ of La$_{0.5}$Ce$_{0.5}$Ni$_{9-x}$Cu$_x$Ge$_4$ normalized per
Ce-mol. The filled symbols represent the La-diluted samples, while the open
symbols represent the pure Ce-compounds CeNi$_{9-x}$Cu$_x$Ge$_4$ for identical Cu-contents.}
\end{center}
\end{figure}
La$_{0.5}$Ce$_{0.5}$Ni$_{9-x}$Cu$_x$Ge$_4$ were prepared by arc
melting the pure elements under argon atmosphere. Subsequently the
samples were annealed at 950$^{\circ}$\,C for two weeks in
evacuated quartz tubes. Less than 0.8\% weight loss occurred during
the melting process. X-ray powder diffraction, optical emission spectroscopy
in an indicatively coupled plasma (ICP-OES) and energy dispersive X-ray spectroscopy (EDX)
indicated that the samples display a single phase character. The system
crystalizes in the tetragonal LaFe$_9$Si$_4$-type structure (space group \emph{I}4/$mcm$). For details of the
preparation and the measurements on the pure Ce-compounds
CeNi$_{9-x}$Cu$_x$Ge$_4$ see \cite{Peyker2009}.\\
Figure\,\ref{fig1} shows the magnetic susceptibility $\chi(T)$ in
the temperature range 2\,K$<$\,$T$\,$<$\,400\,K of
La$_{0.5}$Ce$_{0.5}$Ni$_{9-x}$Cu$_x$Ge$_4$ normalized per Ce-mol
(filled symbols) in comparison with the pure Ce-alloys (open
symbols) taken from \cite{Peyker2009}. For a direct comparison of
the magnetically dilute solid solution
$\mathrm{La}_{0.5}\mathrm{Ce}_{0.5}\mathrm{Ni}_{9-x}\mathrm{Cu}_x\mathrm{Ge}_4$
with the corresponding solid solution
$\mathrm{CeNi}_{9-x}\mathrm{Cu}_x\mathrm{Ge}_4$ with undiluted
Ce-sublattice, the specific heat and magnetic susceptibility data
are normalized to the Ce-content. In case of the Ce-normalized
magnetic susceptibility ($\chi_{\mathrm{Ce-mol}}$), we first
subtracted the nonparamagnetic La-contribution and then scaled
the data with the Ce-concentration, using the following equation:
\begin{equation}\label{eq.norm}
\chi_{\mathrm{Ce-mol}}=2\cdot\biggl(\chi(\mathrm{La}_{0.5}\mathrm{Ce}_{0.5}\mathrm{Ni}_{9-x}\mathrm{Cu}_x\mathrm{Ge}_4)
-0.5\cdot\chi(\mathrm{La}\mathrm{Ni}_{9}\mathrm{Ge}_4)\biggr)
\end{equation}.
For $T>80$\,K, $\chi(T)$  follows a modified Curie-Weiss law,
$\chi(T)=C/(T-\Theta)+\chi_0$, yielding an effective magnetic moment
of $\mu_\mathrm{eff}\approx2.5\mu_\mathrm{B}$, as theoretically
expected for a Ce$^{3+}$-lattice. In the low temperature range
($T<80$\,K) the data scales for $x\leq0.4$ with the Ce-concentration
indicating single-ion behavior. For $x>0.4$ the single-ion character
vanishes and the temperature dependence of the La-substituted
samples deviate from the behavior of the pure Ce-compounds, which
follow a Curie-Weiss-law only down to 100\,K, due to the formation
of an antiferromagnetic transition at lower temperatures
\cite{Peyker2009}. The different temperature dependence of $\chi
(T)$ is due to the absence of antiferromagnetic correlations in the
La-substituted system and results from the
dilution of the magnetic moments.\\
The specific heat $C$ normalized per Ce-mol and divided by temperature $T$ is displayed in Fig.\,\ref{fig2}
for La$_{0.5}$Ce$_{0.5}$Ni$_{9-x}$Cu$_x$Ge$_4$ in the
temperature range between 0.05\,K and $T$\,$<$\,300\,K.
\begin{figure}
\begin{center}
\includegraphics[width=22pc]{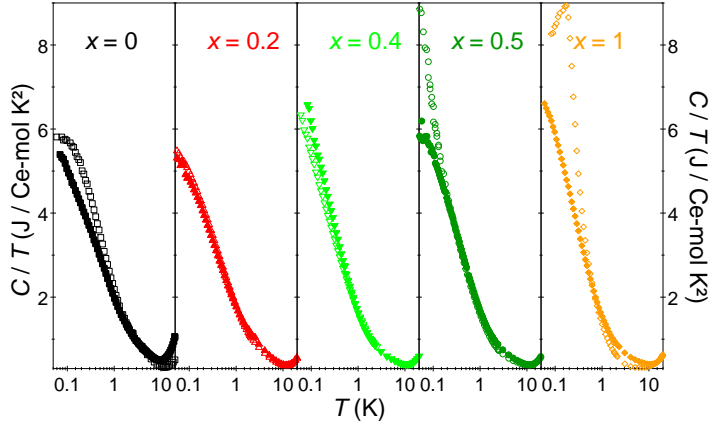}
\caption{\label{fig2}(Color online) The specific heat $C$ divided by
temperature $T$ of La$_{0.5}$Ce$_{0.5}$Ni$_{9-x}$Cu$_x$Ge$_4$
normalized per Ce-mol (filled symbols) and of the undiluted
Ce-compounds CeNi$_{9-x}$Cu$_x$Ge$_4$ (open symbols).}
\end{center}
\end{figure}
As already known from literature \cite{Killer2004} a normalization
to CeNi$_9$Ge$_4$ for $x=0$ is not possible due to the coherence of
the Kondo-lattice. This is, however, not true for the diluted
sample, where a logarithmic increase of $C/T$ below 1.5\,K is
observed, indicating non-Fermi-liquid-behavior. For $0<x\leq0.4$ the
data scale with the undiluted compounds which is in agreement with
single-ion effects as already observed in the magnetic
susceptibility. The absence of the antiferromagnetic transition for
$x>0.4$ in the Ce-diluted compounds is also in line with their
susceptibility behavior. The stronger increase of $C/T$ of the pure
Ce-alloys compared to the diluted compounds is due to the additional
entropy required for the antiferromagnetic ordering.\\

\section{Discussion and conclusions}
\begin{figure}
\begin{center}
\includegraphics[width=30pc]{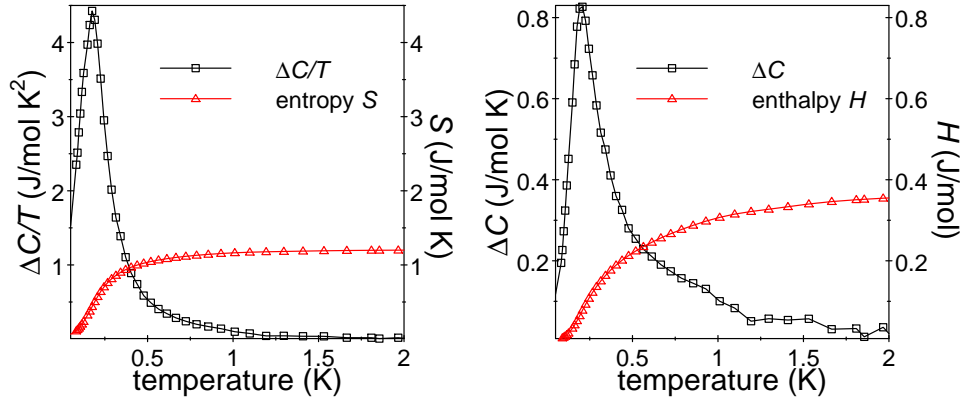}
\caption{\label{fig3}(Color online) The differences of $C/T$ (left panel) and $C$
(right panel) of CeNi$_8$CuGe$_4$ and La$_{0.5}$Ce$_{0.5}$Ni$_8$CuGe$_4$ and
the resulting entropy $S$ and
enthalpy $H$, respectively.}
\end{center}
\end{figure}

In order to study the antiferromagnetic transition of
CeNi$_8$CuGe$_4$ in a more quantitative manner, we take a closer
look at the $C/T$-difference of the diluted and undiluted systems.
Due to the fact, that only the pure Ce-compounds order
antiferromagnetically, the difference in the specific heat $\Delta
C/T$ provides an entropy, which only belongs to the formation of
long range magnetic order. The left picture in Fig.\,\ref{fig3}
displays $\Delta C/T$, showing the
contribution of the antiferromagnetic ordering, and the associated
entropy $S$ in CeNi$_8$CuGe$_4$. The estimated value of the entropy
$S=1.2$\,J/molK is about 20\% of the theoretical expected value of
$R\ln2\approx 5.8$\,J/molK. This is in line with the presence of a
partially Kondo-screened long range antiferromagnetic order which
has been analyzed in terms of the resonant-level model model by
Schotte and Schotte in Ref.~\cite{Peyker2009}. A model calculation
with a RKKY coupling parameter $J=2.3$\,K and a Kondo temperature
$T_K=1.3$\,K approximately reproduces the reduced magnitude of the
AF specific heat anomaly and the enhanced electronic specific heat
anomaly of CeNi$_8$CuGe$_4$. This calculation implies Kondo-screening
of an ordered Ce-moment along the $c$-axis which reduces
the Ce moment to 36\,\% of its CF ground state value of $\mu_c$.
Considering a reduction of the local symmetry at the Ce-sites due to
substitutional disorder present in CeNi$_8$CuGe$_4$ we expect some
reduction of $\mu_c$ as compared to CeNi$_9$Ge$_4$ with
$\mu_c=2\mu_{\rm B}$ \cite{Michor2006} for the CF ground state. The
Kondo-screened ordered moments of CeNi$_8$CuGe$_4$ are thus expected
to range between 0.5\,--\,0.7\,$\mu_{\rm B}$.\\
Further details can be drawn from the calculation of
the enthalpy $H$. Therefore the difference of the specific heat
$\Delta C$ was integrated as displayed in the right panel in
Fig.\,\ref{fig3}. From the estimated value of $H=0.35$\,J/mol an
internal magnetic field of $B=0.13$\,T is determined, using the relation
$H=N_\mathrm{A}0.5\mu_\mathrm{B}B$ with the reduced magnetic moment
of 0.5\,$\mu_\mathrm{B}$ discussed above. This means that an external magnetic field
of about 0.13\,T should lead to a suppression of the longrange antiferromagnetic
order. With the knowledge of this critical magnetic field an estimation of
the N\'{e}el-temperature $T_\mathrm{N} = 87$\,mK can be made which
is by a factor of two smaller than the observed N\'{e}el-temperature $T_\mathrm{N}\approx175$\,mK of
CeNi$_8$CuGe$_4$ \cite{Peyker2009}. From our thermodynamic considerations,
taking into account the experimental N\'{e}el-temperature, a reduced Ce magnetic
moment of 0.24\,$\mu_\mathrm{B}$ would be expected.
\section{Summary}
Comparative studies of the specific heat and the magnetic
susceptibility on the diluted system
La$_{0.5}$Ce$_{0.5}$Ni$_{9-x}$Cu$_x$Ge$_4$ verify that the behavior
of the pure Ce system CeNi$_{9-x}$Cu$_x$Ge$_4$ in the none ordered magnetic
region ($x\leq0.4$) is driven by a single-ion Kondo-effect.
In the magnetic ordered phase ($x > 0.4$) the Kondo-effect
still influences the magnetic ordering, leading to a reduction of
the magnetic moments and therefore to a reduced antiferromagnetic
contribution in the specific heat at $T_\mathrm{N}$, as it is also
predicted in \cite{Peyker2009}, where a resonant-level model in
combination
with a molecular field approach is used. Thermodynamic calculations support these results.\\
\section{Acknowledgments}
This work was supported by the Deutsche Forschungsgemeinschaft (DFG)
under Contract No. SCHE487/7-1 and by the COST P16 ECOM project of
the European Union.

\providecommand{\newblock}{}

\end{document}